\documentclass{PoS}

\title{IceCube}

\ShortTitle{IceCube}

\author{\speaker{Albrecht Karle}, for the IceCube Collaboration\\
        Department of Physics and Wisconsin IceCube Particle Astrophysics Center, University of Wisconsin-Madison, Madison, WI 53706, USA\\
        E-mail: \email{karle@icecube.wisc.edu}}

\abstract{In May 2011, the IceCube neutrino observatory with one cubic 
kilometer instrumented volume started full operation with 5160 sensors
on 86 strings and 324 sensors on 162 IceTop detectors.
The fine-tuning of operation and calibration of the detector 
is still in progress while a very high uptime of well above 98\% is obtained.
New analysis techniques rely on veto techniques for enhanced rejection of atmospheric 
muon and neutrino backgrounds. 
We will give an overview of recent results including the techniques 
of searching for starting tracks and some  comments on the reported evidence 
of astrophysical neutrinos at energies above 30 TeV. 
}

\FullConference{XV Workshop on Neutrino Telescopes,\\
		11-15 March 2013\\
		Venice, Italy}

\begin{document}

\section{IceCube completion and operation}

\begin{figure}[h]
\begin{center}
	\resizebox{0.7\linewidth}{!}{\includegraphics{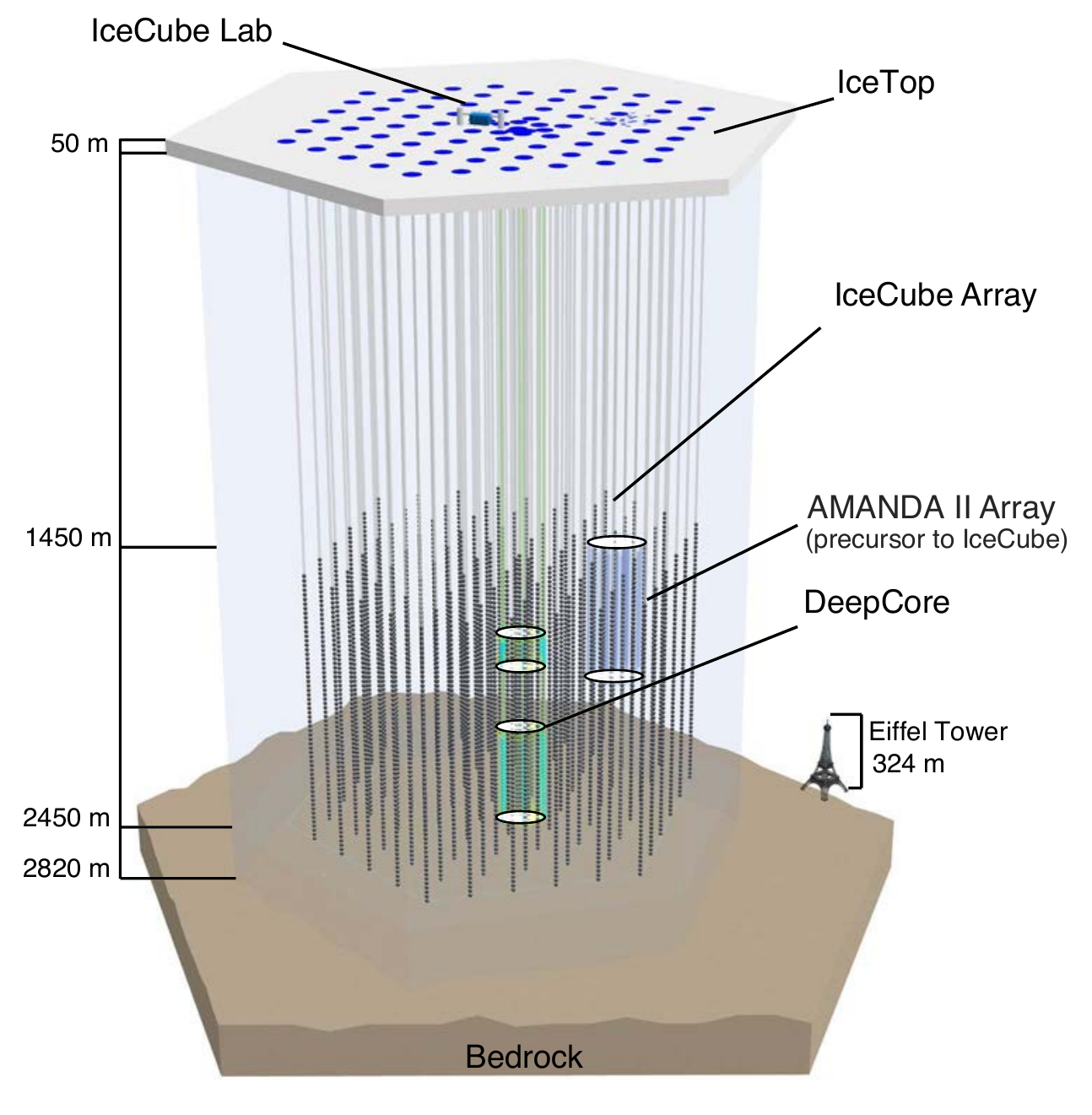}}
	\caption{Schematic view of IceCube.}
	\label{fig:icecube}
\end{center}
\end{figure}

In May 2011, IceCube, a neutrino telescope with one cubic 
kilometer instrumented volume started full operation with 5160 sensors
on 86 strings and 324 sensors in 162 IceTop detectors. 
The plan to build an experiment of this scale was based 
on a decade of research and the demonstration that 
ice was a suitable medium.
First, in the 1990s, the Antarctic Muon and Neutrino Detector Array (AMANDA) was built. 
Then, based on AMANDA as a proof of concept, the 
full kilometer-scale IceCube neutrino telescope was 
constructed and completed by 2010 (see Fig. \ref{fig:icecube}). 
Today, the South Pole has become a premier site for neutrino astronomy.

In the 1990-91 austral summer, the first exploratory effort was 
made at the South Pole to deploy photomultipliers in 
ice at a shallow depth. 
This would be the first of 13 polar seasons that involved 
hot water drilling with the goal of deploying photomultipliers in ice and 
advancing AMANDA and later IceCube.  
It was preceded by an important exploration of the idea to deploy 
PMTs in natural ice in Greenland in 1990.
The result, the 
``Observation of muons using the polar ice cap as a Cerenkov detector"
was published in  \cite{Lowder} and marks an important milestone.  
The authors concluded: ``Our results suggest that a full-scale Antarctic ice 
detector is technically quite feasible," and started making preparations
for an exploration at the South Pole. 
While the conclusion may have sounded far fetched, exactly
20 years later a cubic kilometer detector was indeed in operation.  
Table 1 summarizes the chronological development of string installation and 
some performance measures from AMANDA  to the completion of IceCube.

\begin{table}[th]
\begin{center}
\begin{tabular}{p{2.0cm} p{2.8cm} c c c c c}
\hline

\textbf{Season} & \textbf{Campaign} & \textbf{Sensors} & \textbf{Strings} & \textbf{Depth} & \textbf{Neutrinos} & \textbf{Resol. }\\
      &  & cumul. & season/cum. & (m) & per day & @100TeV \\
 \hline			
1991-1992 & exploratory& few &  & shallow  & - &  \\ 
 \hline					
1992-1993 &  &  &  &  &  & \\
\hline 
1993-1994 & AMANDA-A & 80 & 4 & 800-1000 & - & \\
\hline 
1994-1995 &  &  &  &  &  & \\
 \hline 
1995-1996 & AMANDA-B4 & 86 & 4 & 1500-1950 & $\sim0.01$ & \\
\hline 
1996-1997 & AMANDA-B10 & 206 & 6/10 & 1500-1950 & $\sim1$ & $4\,^{\circ}$\\
\hline 
1997-1998 &  &  &  &  &  & \\
\hline 
1998-1999 & AMANDA-II-13 & 306 & 3/13 & 1500-2300 &  & \\
\hline 
1999-2000 & AMANDA-II-19 & 677 & 6/19 & 1500-1950 & $\sim5$ & $2\,^{\circ}$ \\ 
\hline 
2001-2002 &  &  &  &  &  & \\ 
\hline 
2002-2003 &  &  &  &  &  & \\ 
\hline 
2003-2004 & IceCube prep. &  &  &  &  & \\ 
\hline 
2004-2005 & IceCube 1 & 60 & 1/1 & 1450-2450 & 0.01 & \\ 
\hline 
2005-2006 & IceCube 9 &  540 & 8/9 & 1450-2450 & 2  & \\ 
\hline 
2006-2007 & IceCube 22 & 1320 & 13/22 & 1450-2450 & 18 & $1.5\,^{\circ}$  \\ 
\hline 
2007-2008 & IceCube 40 & 2400 & 18/40 & 1450-2450 & 40 &  $0.8\,^{\circ}$ \\ 
\hline 
2008-2009 & IceCube 59 & 3540 & 19/59 & 1450-2450 & 120 &  $0.6\,^{\circ}$ \\ 
\hline 
2009-2010 & IceCube 79 & 4740 & 20/79 & 1450-2450 &  180 & $0.4\,^{\circ}$ \\ 
\hline 
2010-2011 & IceCube 86 & 5160 & 7/86 & 1450-2450 & \textgreater 200 & $0.4\,^{\circ}$ \\ 
\hline 
\end{tabular}
\end{center}
\caption{The table summarizes the deployment of optical sensors at the 
South Pole.  The cumulative number of sensors deployed per year is shown (324 IceTop sensors deployed with IceCube are not included).  The angular resolution is shown for the reference analysis for point source searches.}
\label{table:table1}
\end{table}

The reliability and successful installation of the sensors was a critical requirement.
About 80 sensors out of 5484 did not commission successfully after the installation. 
The reliability after commissioning has been very high.  
In total 6 sensors failed since the regular science run of the full IceCube detector started in May 2011. 
In October 2013 98.5\% of the deployed sensors are in regular data taking mode. 
Of the 323 IceTop DOMs only one channel failed.  The table 
summarizes the numbers of operational DOMs. 
The failure rates in the last two years since construction has finished 
are very small indeed,  with loss rates at a level of about $0.5\cdot10^{-3}$/year. 

\begin{center}
    \begin{tabular}{p{10.cm} l  p{3cm} c  l }
    \hline
Total number of sensors (DOMs) deployed &  5484  \\ 
DOMs in regular readout at start of full IceCube (May 2011) &  5400 \\ 
DOMs  in regular readout (October 2013)  & 5397   \\  \hline
\end{tabular}
\end{center}

The detector has been running very stably with little downtime. 
During the most recent science run, 5/2012 - 5/2013, the data acquisition 
systems of the detector recorded an uptime of 98.5\%. 
Standard physics analyses use a data set with additional quality criteria, 
which led to a standard physics uptime of about 96.3\%.  
The remaining data could be mined in case of an astronomical event, such as a supernova.  
A science run refers to a fixed detector configuration of typically 1 year during which trigger and filter settings
and any other detector configuration parameters are kept constant or only minor changes 
are performed that will not affect the event selection.  

Efforts are still continuing to optimize the data acquisition and filtering. 
One significant change in January 2013 was the implementation of a system 
that buffers all  photomultiplier signals for several hours on disk. 
Normally only triggered events and scaler rates are recorded. 
The scaler rates (2 ms binning) allow the search for Supernova neutrino bursts. 
A galactic supernova would record a huge burst with time structures 
on a ms scale.  The new system, referred to as Hit-spooling, will allow to extract all single photoelectrons 
from disk up to several hours after a trigger took place.  That means all possible information 
will be kept for secondary fine-tuned analysis beyond the regular online supernova trigger system.
This feature also provides a measure to mitigate against a DAQ crash in 
case of an extremely close (<0.5 kpc) and therefore extremely strong neutrino burst. 
The technology may also be used as a basis for improved extraction 
of long duration events as expected from slowly moving exotic 
particles, for example magnetic monopoles.  

As the statistics of recorded neutrinos increases at a rate of 
more than 50,000 neutrinos/year statistical errors shrink rapidly. 
The optical properties of the ice are known to better than 10\% as a function of depth
\cite{icepaper}.  However, closer inspection has revealed more 
subtle features that have consequences for some analyses. 
This includes the tilt of dust layers of as much as 60\,m vertical variation 
over a horizontal distance of 1\,km. 
Another confirmed feature is the observation that light scatters 
about 10\% less in one horizontal direction than perpendicular to 
that direction. The direction of reduced scattering  coincides
with the direction of the glacial flow, likely not a coincidence. 
The effect is noticeable for high energy cascade event reconstruction. 
In muons, which are triggering IceCube at a rate of 3\,kHz, the moon shadow 
is seen on a monthly basis.  An analysis with a deficit of 8700 events (14$\sigma$)
in one year agrees with predictions \cite{moonpaper}.  The center of the moon shadow 
confirms the absolute pointing to within the error of 0.1$^{\circ}$
after applying a correction of  0.05$^{\circ}$ to account for Earth's  magnetic field.

\begin{figure}
\begin{center}
	\resizebox{0.7\linewidth}{!}{\includegraphics{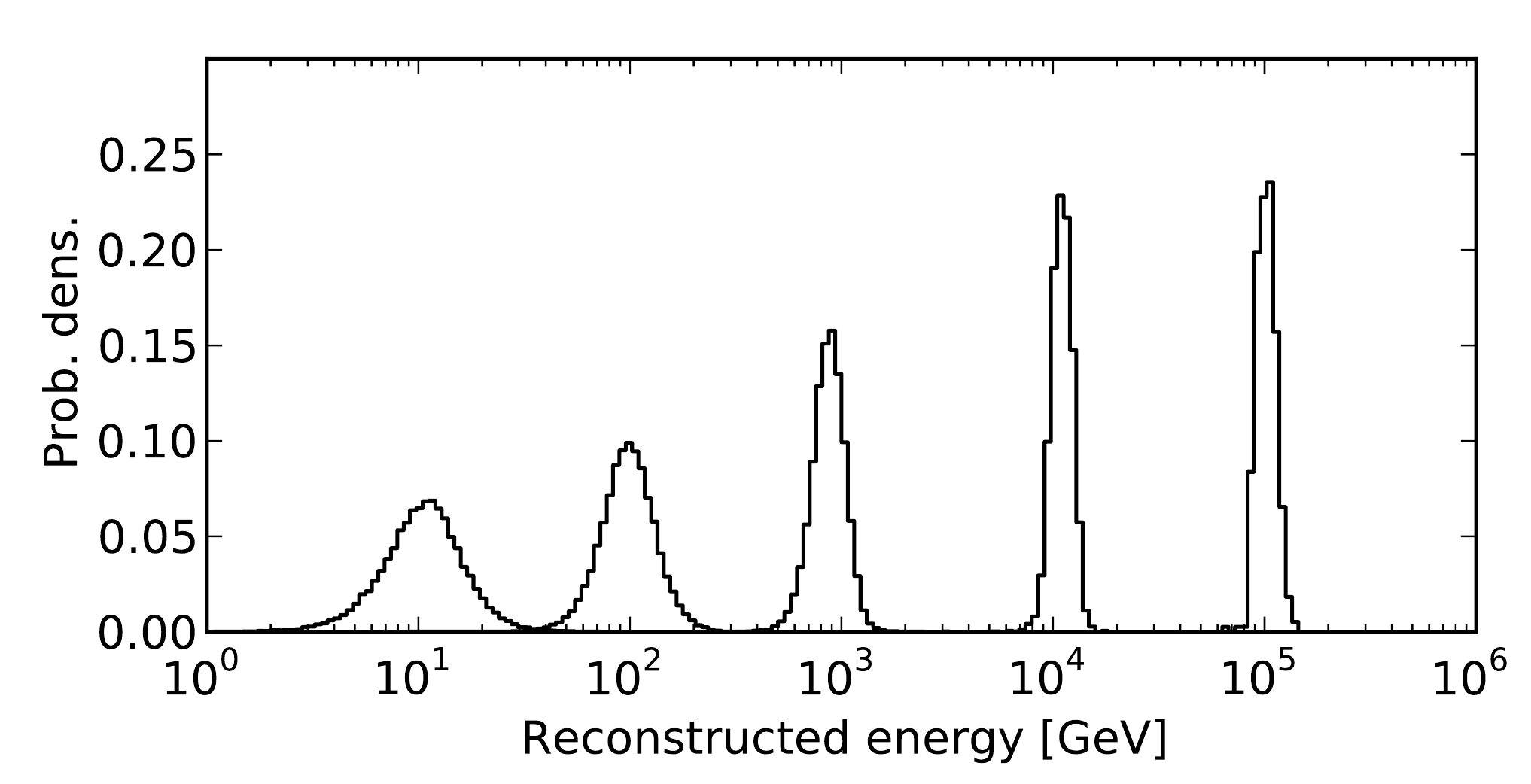}}
	\caption{Reconstructed-energy distribution of simulated cascade events with 
	a fixed true total energy depositions of $10^1, 10^2, 10^3, 10^4$, and $10^5$ GeV.}
	\label{fig:cascade_energy}
\end{center}
\end{figure}

Event reconstruction continues to be an area of ongoing development. 
Single muon event reconstruction results in an angular resolution
at the level of 0.4$^{\circ}$  at energies of 100 TeV.  
We expect significant further improvements for events at this energy and above. 
Figure \ref{fig:cascade_energy} shows the energy resolution of cascades
as a function of energy \cite{energyrecopaper}.  
At energies above 10 TeV the energy resolution for neutrino induced cascades is below 
10\% of the deposited energy.  
The absolute energy calibration is constrained by a variety of methods using both muons
and artificial light flashers to a level of  better than $10$\%.  
Significant progress has been made with unfolding the stochastic energy losses 
of energetic muons above 100\,TeV.  However, even at a sampling length of 1\,km along the track, 
the relation between energy loss and muon energy
remains moderate at the level of 0.3 in log(energy), 
due to the stochastic nature of the energy losses. 
For throughgoing muons the relation to the neutrino energy is further constraint.
These methods however will allow to identify tau neutrino double bang events 
readily at decay lengths of 50\,m and possibly less. 
They also help provide tools to classify single muons in the 
background of high energy muon bundles.  
Equally important they form the basis for further improvements in the angular resolution.

\section{Searches for astrophysical neutrinos} 

\begin{figure}[h]
\begin{center}
	\resizebox{1.0\linewidth}{!}{\includegraphics{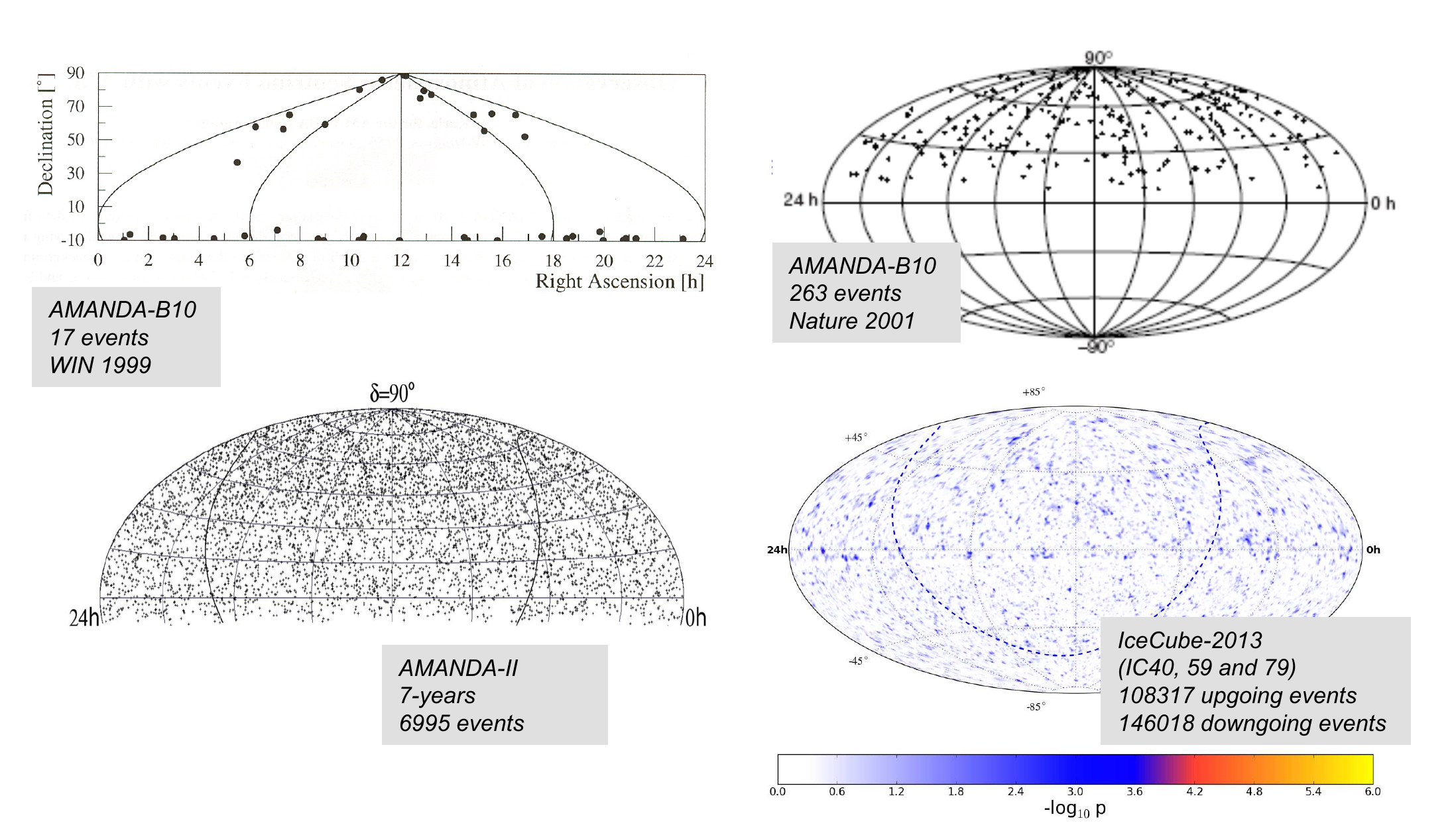}}
	\caption{Neutrino sky maps generated by several generations of neutrino detectors 
	at the South Pole from AMANDA to IceCube.
	 }
\label{fig:skymaps}
\end{center}
\end{figure}

The classical detection channel for neutrino astronomy is muon neutrinos 
with the goal of point source searches because of the superior angular resolution
of muons compared to cascade events.  
To complement the brief construction review in the previous section, we illustrate 
the progress in point source searches over the years in Fig. \ref{fig:skymaps}, 
which includes skymaps from the first map with AMANDA-B10
containing 17 events to the 3 year combined sky map based on 
IceCube data from the 40-, 59- and 79-string configuration. 
The IceCube significance map (IceCube-2013) is based on $1.1\cdot10^5$ 
upgoing neutrinos and 1.46$\cdot10^5$ downgoing events, the latter ones being 
largely background from atmospheric muons.  
The most recent point source results are presented and discussed in a separate 
paper at this conference \cite{JuananVenice} and in \cite{pointsource2013}.
The Southern sky searches are primarily sensitive to sources
above PeV energies because of the much higher 
cosmic ray muon background. The Northern sky searches are effective at all 
energies from TeV to PeV energies where absorption in the Earth begins to attenuate the signal.   
An alternate strategy for rejecting the down going muon background 
relies on an event selection that aims to reject  throughgoing muons altogether. 
Several analyses are underway in IceCube that select events with various degrees
of cosmic ray muon rejection by requiring
that the vertex of down going events or even all events be contained. 

\subsection{Diffuse searches}

\begin{figure}[h]
\begin{center}
	\resizebox{1.0\linewidth}{!}{\includegraphics{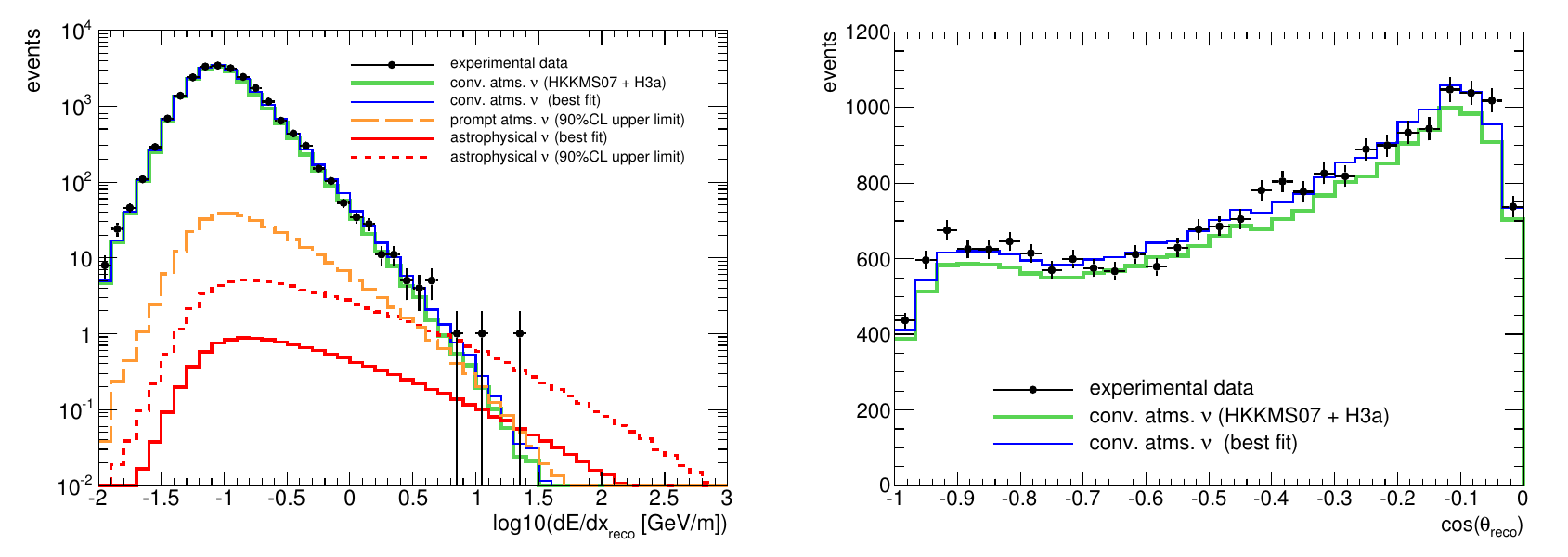}}
	\caption{Search for a diffuse neutrino flux with IceCube's 59 string configuration \cite{ic59diffuse}. 
	The observed energy loss distribution of upgoing muons is compared to
	the best fit of atmospheric and astrophysical neutrino fluxes (left).  The best fit includes a 
	non zero astrophysical component.  The corresponding zenith distribution is shown in the right panel. }
\label{fig:ic59diffuse}
\end{center}
\end{figure}

Astrophysical neutrino fluxes are expected at higher energies than 
the steeply falling spectrum of atmospheric neutrinos. 
Diffuse searches rely on energy only and to some extent on flavor discrimination.  
Background determination is more challenging as it must rely 
to a higher degree on simulations and the modeling 
of the atmospheric neutrino background at high energies.  
Diffuse searches can be performed with $\nu_\mu$ as well as with cascades 
and depending on energy in one or both hemispheres. 

Diffuse searches for astrophysical  $\nu_\mu$-events  in AMANDA \cite{amanda_diffuse}, BAIKAL \cite{baikal} as well as more recently with ANTARES \cite{antares_diffuse}
and an IceCube-40 \cite{ic40diffuse} analysis resulted in upper limits with no indication 
of a hard component.  The flux limits are commonly presented 
as model tests of reference flux with an energy$^{-2}$ spectrum, 
which is motivated by a natural spectral index of 2 of high energy cosmic rays 
generated by shock acceleration. 
The IceCube-59 data set based on 348 days of live time contained 21,943 events 
in the final event selection.  The neutrino purity of the 
strictly upgoing event sample (dominated by atmospheric neutrinos) is 98.8\%. 
Figure \ref{fig:ic59diffuse} shows zenith distribution and the energy loss distribution of data 
compared to atmospheric neutrino backgrounds and an astrophysical flux. 
The final result is in tension with no signal at the 2 sigma level \cite{ic59diffuse} 
and an upper limit is derived. 
All known sources of systematic errors were included into the final fit result.

The cascade detection channel focuses on events with contained vertex 
generated in $\nu_e$ and $\nu_\tau$ charged current interactions and 
neutral current interactions of all neutrino flavors. 
Results from a search for cascade-like high-energy events with the IceCube 40-string detector configuration
 \cite{eike} 
showed an excess of events at a similar level.  
The significance of that excess is $2.7\,\sigma$ with respect to the expectation of conventional atmospheric and prompt atmospheric neutrinos. The upper limit derived from that analysis is an all-flavor flux of $E_{\nu}^2 \Phi (E_{\nu}) = 7.46 \cdot 10^{-8}\,\textrm{GeV}\,\textrm{cm}^{-2}\,\textrm{s}^{-1}\,\textrm{sr}^{-1}$ (90\% confidence level). 
Assuming equal mixing of neutrino flavors when arriving at Earth, that flux is compatible with the best-fit flux and the upper limit derived in the IC59 muon neutrino analysis.

\section{PeV neutrinos and the search for starting tracks}

\begin{figure}
\begin{center}
	\resizebox{0.6\linewidth}{!}{\includegraphics{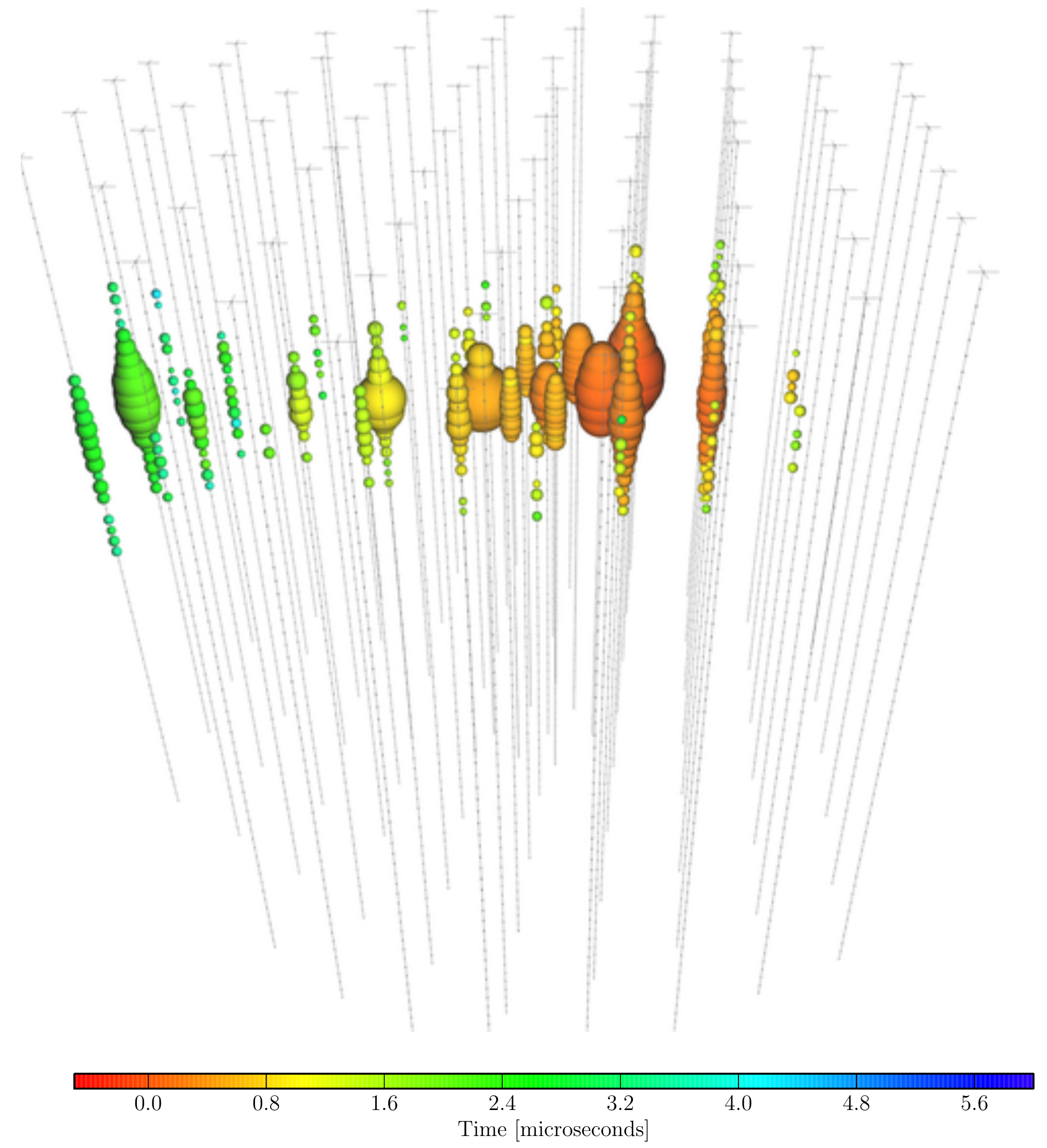}}
	\caption{Horizontal muon neutrino event with a contained vertex 
	and an outgoing track.  This event has a deposited energy of 71\,TeV. 
	Most of the observed events are cascade like events.  }
\end{center}
\label{fig:startingtrack}
\end{figure}

\begin{figure*}[b]
  \begin{center}
  \includegraphics[width=0.48\textwidth]{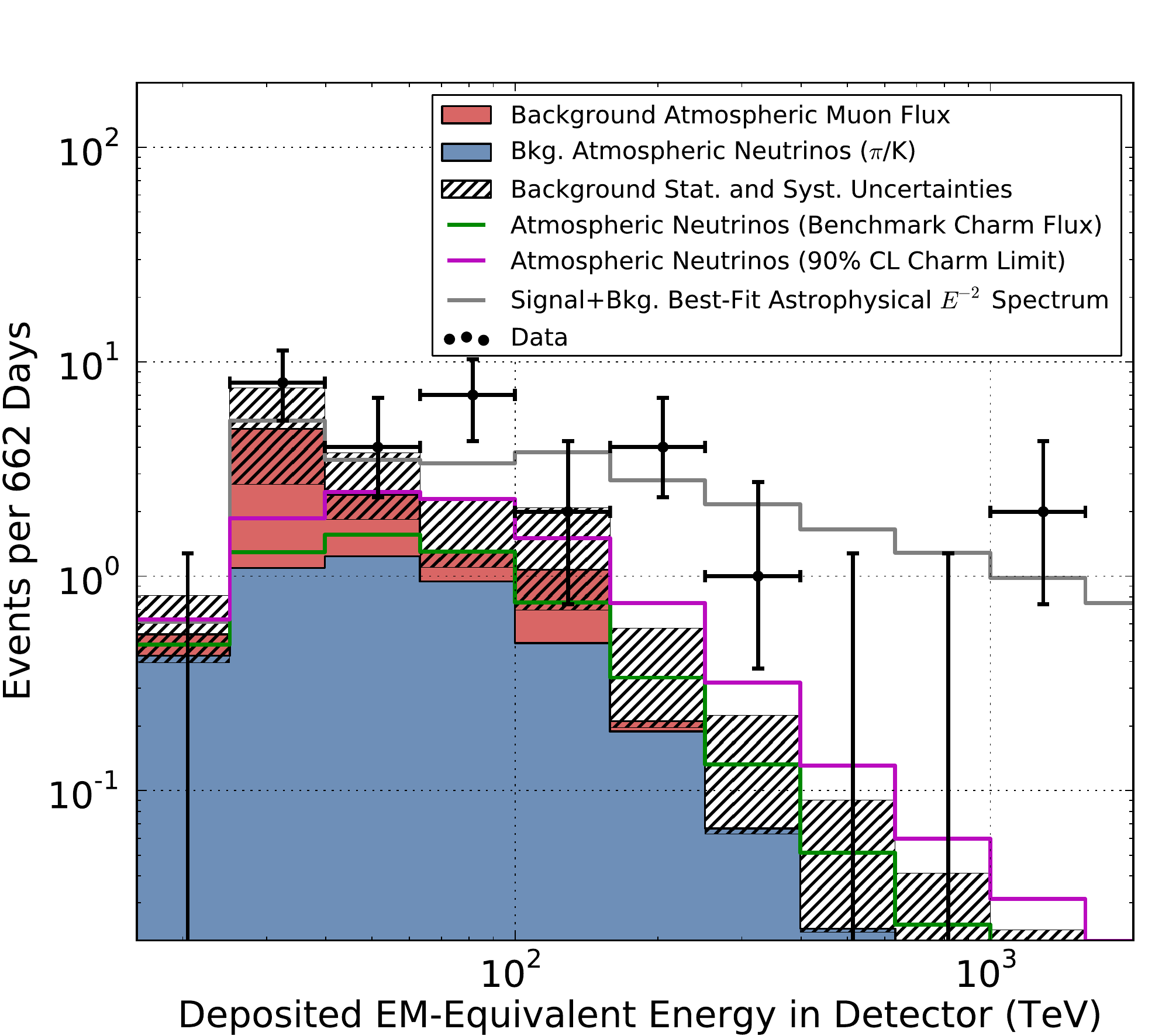}
  \includegraphics[width=0.48\textwidth]{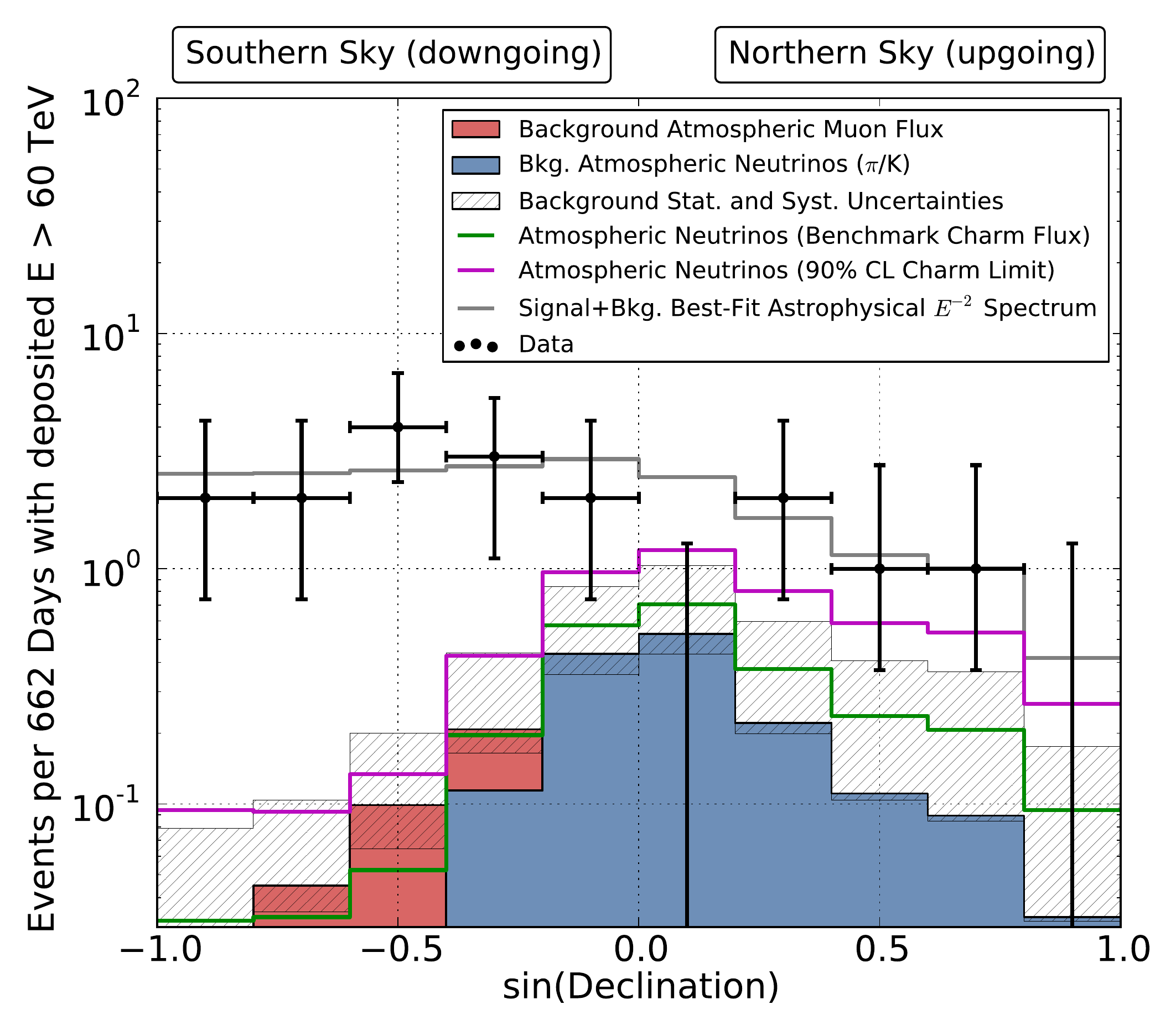}
  \end{center}
  \caption{The distribution of the reconstructed energies of 28 events with contained vertex
  is compared to the best fit for signal and atmospheric background \cite{Sciencepaper}(left).  
  The reconstructed zenith angle distribution for events with reconstructed energy greater than 
  60\,TeV is compared to backgrounds and best fit in the right panel.}
  \label{fig:EnergyAndZenith}
\end{figure*}

IceCube reported the observation of two events the energy of 1 PeV 
above what is generally expected from atmospheric backgrounds and a possible hint 
of an astrophysical source \cite{ehepaper}.
These events were found in a search for cosmogenic or GZK neutrino flux and the two 
events were at the very low end of the energy range that this search was sensitive to.  
In a 
 follow-up analysis a search was performed for neutrinos at lower energies with interaction 
 vertices well contained within the detector volume, discarding events containing muon tracks originating outside of IceCube. 
This event selection achieved nearly full efficiency for interacting neutrinos of all flavors above several hundred TeV, with some sensitivity extending to neutrino energies as low as 30 TeV. 
The event selection relies on relatively simple criteria, essentially requiring that the vertex be well contained 
and rejecting events where early photons were detected inside the veto region consisting primarily of the 
outermost strings and sensors.
An additional 26 events were found for a total of 28 events including the original two PeV events
during a combined live time of 662 days (May 2010 to May 2012). 
The analysis was presented for the first time in May 2013, after this conference, and was recently published in \textit{Science}
\cite{Sciencepaper}. 
Although there is some uncertainty in the expected atmospheric background rates, in particular for the contribution from charmed meson decays, the energy spectrum, zenith distribution, and shower to muon track ratio of the observed events strongly constrain the possibility that these events are entirely of atmospheric origin.
Almost all of the observed excess is in showers which are randomly distributed in the fiducial volume and in direction rather than muon tracks, ruling out an increase in penetrating muon background to the level required.
Figure \ref{fig:startingtrack} shows an event with a reconstructed deposited energy of 71\,TeV.
It is easy to see that this event is indeed an event with the vertex inside. No 
signals have been recorded at all in the outermost layer (on the right side).  
The good energy resolution for contained events and specifically for cascades (Figure \ref{fig:cascade_energy})
is the basis for the energy spectrum shown in Figure \ref{fig:EnergyAndZenith}, 
which shows a significant excess over background at higher energies.

The zenith angle distribution in Fig. \ref{fig:EnergyAndZenith} illustrates the effectiveness of the event selection
and background rejection especially in the downgoing hemisphere.  
In this figure only events with energy above 60\,TeV are shown.  It
can be seen that the atmospheric neutrino background is highly suppressed
for zenith angles less than $\approx 60^\circ$.  
The reason for the suppression of atmospheric neutrinos towards smaller zenith angle 
is the fact that atmospheric neutrinos at sufficiently high energy will be accompanied by 
muons generated in the same parent air shower.  
This mechanism, pointed out in \cite{schoenert_veto}, becomes very effective above energies of order 100 TeV.   
These accompanying muons will trigger the muon veto, removing the majority of these events from the sample and biasing atmospheric neutrinos to the northern hemisphere.
The majority of the observed events, however, arrive from the south and one of the PeV events 
in fact is reconstructed at a zenith angle of only 23$^\circ$ with an angular error of 11$^\circ$.  
A search was performed for clustering of these events, which did not leave any significant 
evidence for a point source in this sample.

\begin{figure}
\begin{center}
	\resizebox{0.9\linewidth}{!}{\includegraphics{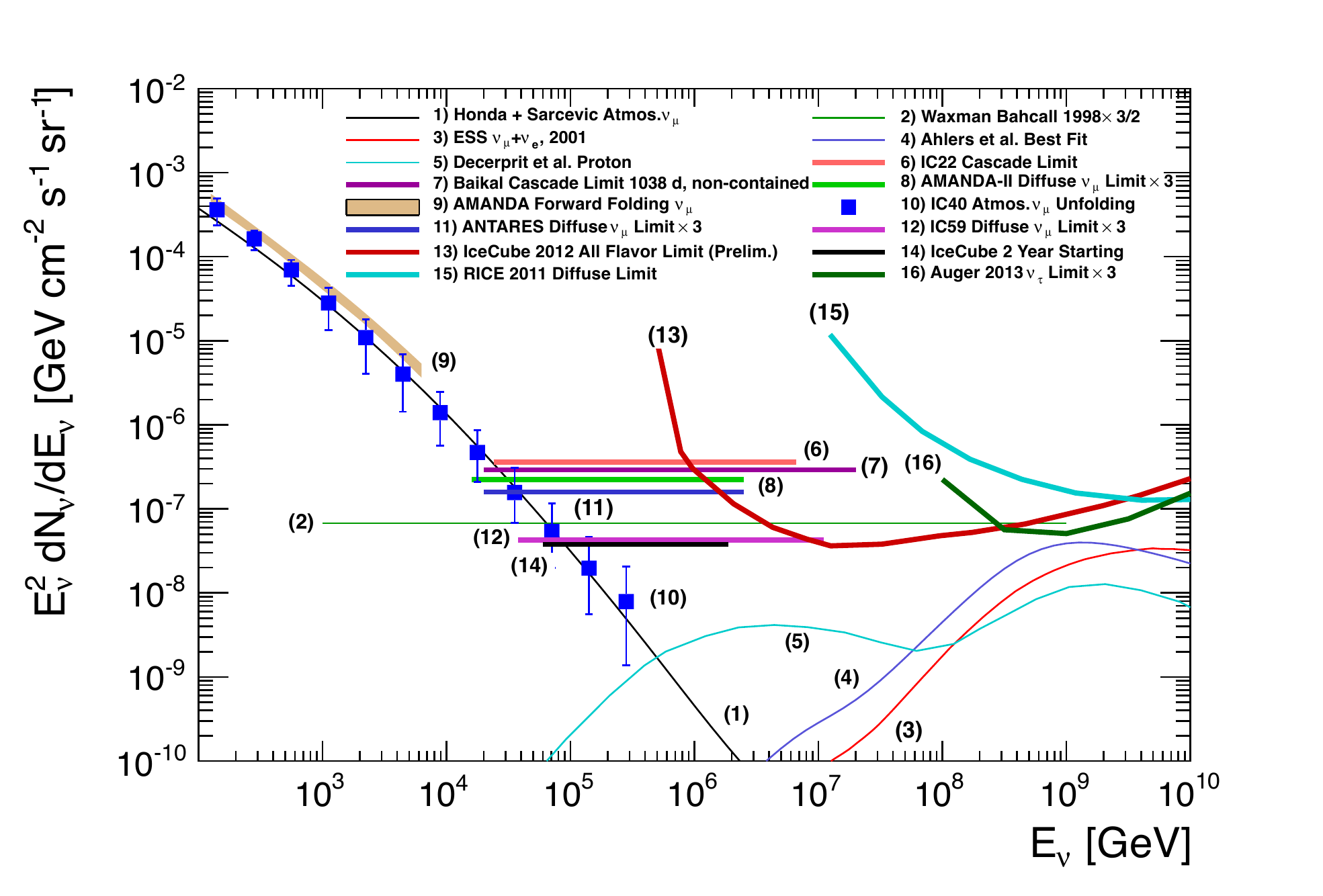}}
	\caption{An overview is presented of observed atmospheric neutrino fluxes, 
	upper limits to diffuse fluxes and models.  The IceCube 2012 differential 
	upper limit (11) turn up sharply at ~1PeV because of observed PeV events. 
	The best fit diffuse flux using starting events in IceCube (12) forms evidence for a diffuse astrophysical flux
	up to PeV energies 
	above the atmospheric neutrino spectrum extending to a few 100 TeV. }
\end{center}
\label{fig:DiffuseFluxes1}
\end{figure}

In a global fit that allows the normalization of the atmospheric neutrino backgrounds to float, 
the data in the energy range between 60\,TeV and 2\,PeV are well described
by an $E^{-2}$ neutrino spectrum with a per-flavor normalization of \\

\vspace{-0.55 cm}
\begin{centering}
$E^2 \Phi(E) = (1.2 \pm 0.4) \cdot 10^{-8}\, \mathrm{GeV}\, \mathrm{cm}^{-2}\, \mathrm{s}^{-1}\, \mathrm{sr}^{-1}$. \\
\end{centering}
The absence of events at higher energies may be an indication for a break of the energy spectrum. 
The result is inconsistent with zero astrophysical flux at the 4 sigma level. 
Figure 7
shows some of the observations  
of high energy neutrino fluxes.  
Atmospheric neutrino fluxes have been measured by IceCube up to 
energies where an indication for a hardening of the energy spectrum is 
observed in several analyses. 
For reference, the atmospheric neutrino flux expectation is shown for Honda \cite{honda}, 
and the upper bound for an astrophysical $E^{-2}$ neutrino spectrum based on the observed energy in high energy cosmic rays derived by  Waxman and Bahcall \cite{waxman}. 
The observed results and upper limits in muon and cascade analyses of IceCube (\cite{Sciencepaper,ic40diffuse,ic59diffuse, EHE_origin} and some earlier results)  
are in reasonable agreement with the first observed evidence at the 4 sigma for an 
astrophysical flux in the starting track analysis \cite{Sciencepaper}.  
Next steps in the diffuse searches will include the inspection of several 
years of diffuse muon neutrinos as well as cascade searches with the full detector. 
The IC59 analysis may be seen as a hint toward an astrophysical muon neutrino flux in the Northern hemisphere 
at a level compatible with the flux reported in the starting track analysis. 
If this is confirmed, the observed diffuse flux can be put in perspective of 
the point source searches already under way since several years.

\section{Concluding remarks} 

The IceCube detector is operating well with  high uptime and exceeding original 
performance parameters.
First evidence is emerging for an astrophysical neutrino flux 
in an analyses relying on events with contained vertex. 
Many other physics results were not discussed in this report. 
A detailed discussion of point source results was presented 
by J. Aguilar \cite{JuananVenice}. 
At low energies intense efforts are underway using the 
DeepCore in fill detector to search for dark matter above 10 GeV and 
determine oscillation parameters using atmospheric neutrinos. 
For these results and an outlook for future upgrades we refer to
the presentation by M. Kowalski at this conference \cite{KowalskiVenice}.

\section{Acknowledgments}

This research was supported in part by the U.S. National Science Foundation-Office of Polar Programs, U.S. National Science Foundation-Physics Division, University of Wisconsin Alumni Research Foundation. 
Thanks  to D. Chirkin, J. Kelley, H. Wissing, M. Kowalski, Ch. Weaver and other collaborators or useful discussions.

\end{document}